\pdfoutput=1
\documentclass[aps,prl,showpacs,preprintnumbers,twocolumn]{revtex4}

\usepackage[colorlinks=true, pdfstartview=FitV, linkcolor=blue, citecolor=blue, urlcolor=blue]{hyperref}
\usepackage[dvipdfmx]{graphicx}
\usepackage{amsmath}
\usepackage{color}

\newcommand{\tr}{\mathrm{tr}}

\begin{document}

\preprint{RIKEN-QHP-113}
\title{Lattice QCD with mismatched Fermi surfaces}

\author{Arata~Yamamoto}
\affiliation{
Theoretical Research Division, Nishina Center, RIKEN, Saitama 351-0198, Japan\\
}

\date{\today}

\begin{abstract}
We study two flavor fermions with mismatched chemical potentials in quenched lattice QCD.
We first consider a large isospin chemical potential, where a charged pion is condensed, and then introduce a small mismatch between the chemical potentials of the up quark and the anti-down quark.
We find that the homogeneous pion condensate is destroyed by the mismatch of the chemical potentials.
We also find that the two-point correlation function shows spatial oscillation, which indicates an inhomogeneous ground state, although it is not massless but massive in the present simulation setup.
\end{abstract}


\maketitle

\paragraph{Introduction.}

Multi-component mixture of quantum systems, in particular, of fermions, exhibits interesting physics.
One of the most famous phenomena is inhomogeneous superconductivity.
In a homogeneous superconductor, electrons with opposite spins and opposite momenta are paired on the Fermi surfaces, and thus the Cooper pair has zero total momentum.
When the populations of spin-up and spin-down electrons are imbalanced, the Cooper pair has nonzero total momentum.
If the population imbalance is large enough to destroy the homogeneity, a superconductor becomes inhomogeneous.
The expectation value of the Cooper pair oscillates spatially.
This inhomogeneous superconductor is called the Fulde-Ferrell-Larkin-Ovchinnikov (FFLO) state \cite{Fulde:1964zz}.
The FFLO state has been also studied in spin-imbalanced cold-atomic systems \cite{Radzihovsky:2010}.

In quantum chromodynamics (QCD), inhomogeneous phases are expected in high quark number density \cite{Rajagopal:2000wf}.
Quarks have many degrees of freedom, such as flavor, color, and chirality.
If the Fermi surfaces of these species are mismatched (because of charge neutrality, mass difference, etc.), the FFLO state can take place.
Inhomogeneous states were predicted in the chiral condensate \cite{Deryagin:1992rw}, the pion condensate \cite{Son:2000xc}, and the diquark condensate \cite{Alford:2000ze}.
Although these inhomogeneous states have been studied theoretically in effective models, it is still unknown whether they really exist or not in QCD, in particular, in the strong coupling regime.
To establish the existence of inhomogeneous states without model ambiguity, the lattice QCD simulation is necessary.

\begin{figure}[h]
\includegraphics[scale=1]{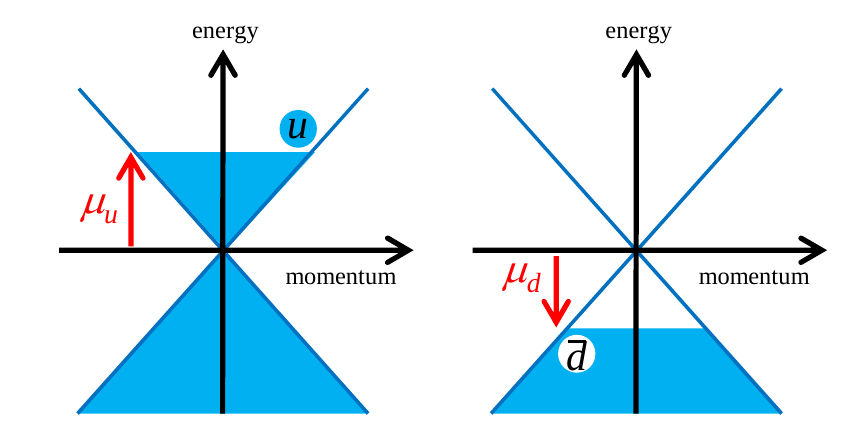}
\caption{\label{fig1}
Mismatched Fermi surfaces.
The massless dispersion relation is depicted for simplicity.
}
\end{figure}

We consider two-flavor fermions with the same mass and different chemical potentials.
We take $\mu_u > 0$ and $\mu_d < 0$.
The schematic figure of the Fermi surfaces is shown in Fig.~\ref{fig1}.
The mismatch between the chemical potentials of the up quark and the anti-down quark is
\begin{equation}
\Delta \mu = \mu_u - |\mu_d|.
\end{equation}
At $\Delta \mu = 0$, when chemical potentials exceed a critical value, the homogeneous condensate of a charged pion is formed,
\begin{equation}
\langle \bar{d}(x) \gamma_5 u(x) \rangle = C,
\label{eqPION}
\end{equation}
where $C$ is a nonzero constant.
At $\Delta \mu \ne 0$, the homogeneous condensate becomes unstable and the FFLO state can be formed as
\begin{equation}
\langle \bar{d}(x) \gamma_5 u(x) \rangle = C\exp(i\Delta\mu x).
\label{eqFFLO}
\end{equation}
The condensate spatially oscillates with the frequency $\Delta \mu$.
There are technical advantages to consider the pion condensate:
(i) the pion condensate is gauge invariant unlike the diquark condensate, (ii) the pion condensate is zero at zero chemical potentials unlike the chiral condensate, and (iii) the pion correlation function is the least noisy in the lattice QCD simulation.

In this Letter, we study lattice QCD with mismatched chemical potentials.
In the lattice QCD simulation with finite chemical potentials, there is the sign problem.
In this study, we adopt the quenched QCD simulation where the contribution from the fermion determinant is neglected in gauge ensembles.
Calculating the correlation function of a charged pion, we discuss how the mismatched chemical potential affects the homogeneous state and the inhomogeneous state.

\paragraph{Homogeneous analysis.}

We performed the quenched QCD simulation with the plaquette gauge action and the Wilson fermion action.
The bare lattice coupling is $\beta = 6/g^2 = 5.9$, where the lattice spacing is $a\simeq 0.10$ fm.
The hopping parameter is $\kappa_u = \kappa_d = 0.1566$, where the pion mass is $am_\pi \simeq 0.4$.
We used the lattice volumes $V=N_xN_yN_zN_\tau =10^4$, $12^4$, and $14^4$ and perform linear extrapolation to $V\to \infty$.
Spatial boundary conditions are periodic, and a temporal boundary condition is antiperiodic (periodic) for fermions (gluons).

The expectation value of any symmetry breaking condensate vanishes in a finite volume.
To extract the information of the condensate, we calculated the two-point correlation function
\begin{equation}
\begin{split}
& \Pi (x_1-x_2) \\
& \equiv \frac{1}{N_yN_zN_\tau}\sum_{yz\tau} \langle [\bar{d}(x_1) \gamma_5 u(x_1)]^\dagger \bar{d}(x_2) \gamma_5 u(x_2) \rangle \\
& = - \frac{1}{N_yN_zN_\tau}\sum_{yz\tau} \langle \tr \gamma_5 S_{x_1,x_2}(\mu_d) \gamma_5 S_{x_2,x_1}(\mu_u) \rangle.
\label{eqODLRO}
\end{split}
\end{equation}
The quark propagator $S(\mu)$ is given as the inverse of the Dirac operator, $S(\mu)= D^{-1}(\mu)$.
The summation in Eq.~(\ref{eqODLRO}) is performed for the zero-momentum projection of $p_y=p_z=p_\tau=0$.
Since excited states are massive and heavy, the long-range behavior is dominated by the pion condensate,
\begin{equation}
\lim_{x \to \infty} \Pi (x) = |C|^2.
\end{equation}
This is called the off-diagonal long-range order (ODLRO).
In numerical simulations, since the lattice volume is finite, we calculated the correlation function at the largest separation $x = aN_x/2$, and then numerically extrapolate them to the infinite separation limit,
\begin{equation}
\Pi_{\rm ODLRO} \equiv \lim_{N_x \to \infty} \Pi (aN_x/2).
\end{equation}

\begin{figure}[h]
\includegraphics[scale=1.4]{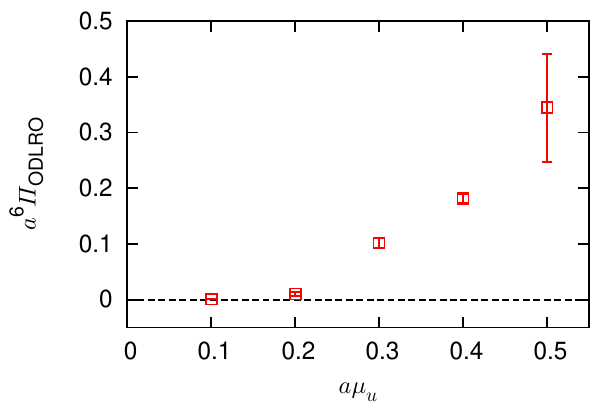}
\caption{\label{fig2}
The ODLRO of the pion two-point correlation function $\Pi(x)$ with $\Delta \mu = 0$.
}
\end{figure}

In Fig.~\ref{fig2}, we plot the ODLRO of the correlation function with $\Delta \mu = 0$.
Above the threshold of a half of the pion mass $am_\pi/2 \simeq 0.2$, the (matched) Fermi surfaces are formed and the ODLRO increases.
This corresponds to the charged pion condensation (\ref{eqPION}) at finite isospin chemical potentials \cite{Kogut:2002tm}.
A charged pion condensate exhibits a homogeneous superconductor.

\begin{figure}[h]
\includegraphics[scale=1.4]{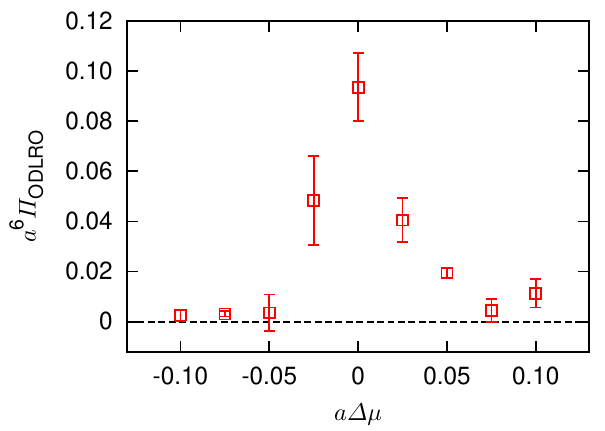}
\caption{\label{fig3}
The ODLRO of the pion two-point correlation function $\Pi(x)$ with $\Delta \mu \ne 0$.
}
\end{figure}

In Fig.~\ref{fig3}, we plot the ODLRO of the correlation function as a function of $\Delta \mu$.
To reduce the computational cost, we fixed the chemical potential of the anti-down quark at $a\mu_d = -0.3$ and varied only the chemical potential of the up quark.
(Thus the data is not symmetric about the transformation $\Delta \mu \leftrightarrow - \Delta \mu$.)
At $\Delta \mu = 0$, the pion condensate is finite.
As $|\Delta \mu|$ increases, the pion condensate decreases.
Eventually, the pion condensate disappears above $|a\Delta \mu| \simeq 0.05$-0.10.
The formation of the Cooper pairs on the mismatched Fermi surfaces is energetically disfavored.

\paragraph{Inhomogeneous analysis.}

In the above analysis, we assumed that the ground state is homogeneous.
However, when the Fermi surfaces are mismatched, the ground state can be inhomogeneous.
If the inhomogeneous pion condensate is formed as Eq.~(\ref{eqFFLO}), the two-point correlation function behaves as
\begin{equation}
\lim_{x \to \infty} \Pi (x) = |C|^2 \cos (\Delta \mu x).
\end{equation}
The expectation value of the imaginary part proportional to $i\sin (\Delta \mu x)$ is exactly zero because the correlation function is symmetric under the parity transformation $x \leftrightarrow -x$.
The long-range behavior is dominated by the lightest state among all possible (homogeneous and inhomogeneous) states.
We can determine whether the ground state is homogeneous or inhomogeneous without assuming it.
If there are degenerate lightest states, e.g., both homogeneous and inhomogeneous condensates, the long-range behavior is given as the sum of them.

In a finite volume, momentum is discretized as
\begin{equation}
p_x = \frac{2\pi}{N_xa}k_x \quad (k_x = 0,1,2, \cdots, N_x-1).
\label{eqP}
\end{equation}
Since the frequency of spatial oscillation satisfies this condition, $\Delta \mu$ should be set to satisfy it.
We used the elongated lattice volume $V=N_x \times N_yN_zN_\tau =50 \times 10^3$.
The minimum momentum unit is $2\pi/N_x \simeq 0.126$.
If one wants smaller momentum unit, one need larger lattice size.
Other simulation parameters are the same as in the homogeneous analysis.

\begin{figure}[h]
\includegraphics[scale=1.4]{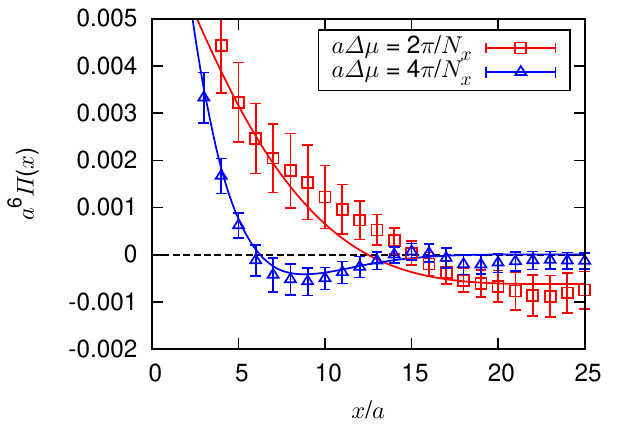}
\caption{\label{fig4}
The pion two-point correlation function $\Pi(x)$.
}
\end{figure}

In Fig.~\ref{fig4}, we show the correlation function with $a\Delta \mu = 2\pi/N_x$ and $4\pi/N_x$.
The chemical potential of the anti-down quark is fixed at $a\mu_d = -0.3$ and the chemical potential of the up quark is set at $a\mu_u \simeq 0.426$ and $0.551$, respectively.
At long range, only the ground-state contribution survives and the excited-state contributions die out.
Since spatial boundary conditions are periodic, the correlation function is reflection symmetric about $x/a=N_x/2=25$.
The ground-state contribution can be fitted as
\begin{equation}
\Pi (x) =  |C|^2 \cosh(-m_\pi(x - aN_x/2)) \cos (\Delta \mu x).
\end{equation}
Although the ground state shows spatial oscillation with the frequency $\Delta \mu$, it is massive.
The best-fit functions with the fit range $x/a = [5,25]$ are shown in Fig.~\ref{fig4}.
The best-fit parameters are $am_\pi = 0.13 \pm 0.02$ for $a\Delta \mu = 2\pi/N_x$ and $am_\pi = 0.33 \pm 0.05$ for $a\Delta \mu = 4\pi/N_x$.
Thus, the ground state is inhomogeneous but does not seem a condensate.

In a finite volume, there is no spontaneous symmetry breaking and the Nambu-Goldstone boson cannot be massless.
If a condensate exists, the mass of the would-be Nambu-Goldstone boson must approach zero, $m_\pi \to 0$, in the infinite-volume limit.
We must carefully take the infinite-volume extrapolation of numerical data.
As far as we searched in this study, we find no tendency of $m_\pi \to 0$ and thus no hints for an inhomogeneous condensate in the infinite-volume limit.
By examining more sets of simulation parameters, we will find the inhomogeneous condensate if it exists in QCD.

\paragraph{Discussion.}

We have studied lattice QCD with mismatched chemical potentials.
We have found that the homogeneous condensate is destroyed by the mismatch of chemical potentials.
Although a two-point correlation function shows oscillating behavior, we have not found the evidence of inhomogeneous condensates in the present simulation parameters.
In condensed matter and atomic physics, it is known that the FFLO phase is quite unstable and it exists only in a narrow parameter region \cite{Radzihovsky:2010}.
Also in QCD, we may need the computational effort for fine tuning of simulation parameters.
In particular, since the discrete momentum (\ref{eqP}) depends on the spatial size, it is troublesome to take the infinite-volume extrapolation with other parameters fixed. 
It might be easier to analyze in two or pseudo-one dimension, because the computational cost is much lower and inhomogeneous phases tend to be favored in lower dimensions.

In this paper, we have adopted the ODLRO of a two-point correlation function to extract the information of the ground state.
An alternative way is to calculate the condensate (i.e., a one-point function) by introducing an explicitly symmetry breaking term $\varepsilon \cos (\Delta \mu x)$.
In this case, we must take $V \to \infty$ and then $\varepsilon \to 0$.
(The order of these limits is crucial.
If $\varepsilon \to 0$ is taken before $V \to \infty$, the condensate is zero.)
In the numerical simulation, since $\varepsilon$ and $V$ are finite, we must extrapolate to $V \to \infty$ and $\varepsilon \to 0$, keeping the spatial size larger than the Compton wavelength of the would-be Nambu-Goldstone boson mass.

We have performed the quenched QCD simulation.
However, we cannot a priori validate the quenched approximation at high density.
The full QCD simulation is necessary for a rigorous analysis, but it is difficult to implement mismatched chemical potentials in full QCD.
In the absence of the mismatch ($\mu_u = -\mu_d$), the full QCD simulation is possible because the total fermion determinant is semi-positive, as
\begin{equation}
\begin{split}
\det \mathcal{D}
& = \det D(\mu_u) \det D(-\mu_u) \\
&= |\det D(\mu_u)|^2 \ge 0,
\end{split}
\end{equation}
where the relation $\gamma_5 D(\mu) \gamma_5 = D^\dagger(-\mu)$ is used.
In the presence of the mismatch ($\mu_u \ne -\mu_d$), the semi-positivity is lost and there is the sign problem.
To avoid the sign problem, we need to consider ideal situations.
For example, in the case of four flavors with two isospin chemical potentials, the fermion determinant is semi-positive, as
\begin{equation}
\begin{split}
&\det \mathcal{D}\\
& = \det D(\mu_u) \det D(\mu_d) \det D(-\mu_u) \det D(-\mu_d) \\
& = |\det D(\mu_u) \det D(\mu_d)|^2 \ge 0.
\end{split}
\end{equation}
The mismatch of the Fermi surfaces exists but the sign problem does not exist.
The same strategy is possible in four flavors with two quark chemical potentials in two-color QCD or two flavors with two chiral chemical potentials.

The author thanks Tomoya Hayata and Yoshimasa Hidaka for useful discussions.
The author is supported by the Special Postdoctoral Research Program of RIKEN.
The numerical simulations were performed by using the RIKEN Integrated Cluster of Clusters (RICC) facility.

\end{document}